\documentclass{IEEEtran}	
\IEEEoverridecommandlockouts

\usepackage{xspace}
\usepackage[T1]{fontenc}

\usepackage{algorithm, algorithmic}
\floatname{algorithm}{Algorithm}

\newfloat{Algorithm}{htpb}{alg}
\newcounter{prgline}
\newcommand{\pl}{\theprgline\addtocounter{prgline}{1}}

\parskip=\smallskipamount

\usepackage[scaled=0.92]{helvet}

\usepackage[latin1]{inputenc}
\usepackage{pdfsync}
\usepackage{url}
\usepackage{graphicx}
\usepackage{amsfonts,amsmath,mathrsfs,theorem,bm}
\usepackage{rotating}
\usepackage{algorithm, algorithmic}
\floatname{algorithm}{Algorithm}
\usepackage[usenames,dvipsnames]{color}
\usepackage{multirow}
\usepackage{dcolumn}
\usepackage{colortbl}
\usepackage{mathrsfs}

\newtheorem{theorem}{Theorem}
\newcounter{noqed}
\newcommand{\qed}{ \ifmmode\mbox{ }\fi\rule[-.05em]{.3em}{.7em}\setcounter{noqed}{0}}
\newenvironment{proof}[1][{}]{\noindent{\bf Proof#1. }\setcounter{noqed}{1}}{\ifnum\value{noqed}=1\qed\fi\par\medskip}

\newcommand{\BB}[1]{{\mathscr B}_{#1}}
\newcommand{\WW}[1]{{\mathscr W}_{#1}}
\newcommand{\hBB}[1]{{\hat{\mathscr B}}_{#1}}
\newcommand{\RR}{{\mathscr R}}

\newcommand{\DD}{\mathscr{D}}

\renewcommand{\epsilon}{\varepsilon}

\renewcommand{\theta}{\vartheta}

\def\..{\,\mathpunct{\ldotp\ldotp}} % Middle stuff for intervals. Usage: \..

\newcommand{\comment}[1]{}

\newcommand{\BEGIN}{\text{\textbf{begin}}\xspace}

\newcommand{\FOREACH}{\text{\textbf{foreach}}\xspace}

\newcommand{\END}{\text{\textbf{end}}\xspace}
\newcommand{\DO}{\text{\textbf{do}}\xspace}

\newcommand{\RETURN}{\text{\textbf{return}}\xspace}

\newcommand{\UNTIL}{\text{\textbf{until}}\xspace}

\newcommand{\FUNCTION}{\text{\textbf{function}}\xspace}

\newcommand{\COMMENT}{\xspace}

\title{In-Core Computation of Geometric Centralities with HyperBall: \\A Hundred Billion Nodes and Beyond} 

\author{\IEEEauthorblockN{Paolo Boldi \qquad Sebastiano
Vigna}\\\IEEEauthorblockA{Dipartimento di Informatica, Universit\`a degli Studi
di Milano, Italy}\thanks{The authors have been supported by the EU-FET grant NADINE (GA 288956).}}

\begin{document}
\bibliographystyle{IEEETran}
\maketitle

\begin{abstract}
Given a social network, which of its nodes are more central? This question 
was asked many times in sociology, psychology and computer science, and a
whole plethora of \emph{centrality measures} (a.k.a.~\emph{centrality indices}, or \emph{rankings}) were 
proposed to account for
the importance of the nodes of a network. 
In this paper, we approach the problem of computing \emph{geometric centralities},
such as \emph{closeness}~\cite{BavMMGS} and \emph{harmonic centrality}~\cite{BoVAC}, on
very large graphs; traditionally this task requires an all-pairs shortest-path computation in the exact case, 
or a number of breadth-first traversals
for approximated computations, but these techniques yield very weak statistical
guarantees on highly disconnected graphs. We rather assume that the graph is accessed in a \emph{semi-streaming}
fashion, that is, that adjacency lists are scanned almost sequentially,
and that a very small amount of memory (in the order of a dozen bytes) per node is available
in core memory. We leverage the newly discovered algorithms based on HyperLogLog
counters~\cite{BRVH}, making it possible to approximate a number of
geometric centralities at a very high speed and with high accuracy. 
While the application of similar algorithms for the approximation of 
closeness was attempted in the MapReduce~\cite{DeGMR} framework~\cite{KPSCLNAO}, our exploitation
of HyperLogLog counters reduces exponentially the memory footprint, paving the way
for in-core processing of networks with a hundred billion nodes using ``just" 2\,TiB of RAM.
Moreover, the computations we describe are inherently parallelizable, and scale linearly with
the number of available cores.  
\end{abstract}

\section{Introduction}

In the last years, there has been an ever-increasing research activity in the
study of real-world complex networks.
These networks, typically generated directly or indirectly by human activity and
interaction, appear in a large variety of contexts and often exhibit a
surprisingly similar structure.

One of the most important notions that researchers have been trying to capture
in such networks is ``node centrality'':
ideally, every node (often representing an individual) has some degree of
influence or importance within the social domain under consideration, and one
expects such importance to be reflected in the structure of the social network.
Centrality in fact has a long history in the context of social sciences:
starting from the late 1940s~\cite{BavMMGS} the problem of singling out
influential individuals in a social group has been a holy grail that
sociologists have been trying to capture for many decades.

Among the types of centrality that have been considered in the literature
(see~\cite{BorCNF} for a good survey), many have to do with the distance to
other nodes. If, for instance, the sum of distances to all other nodes
is large, the node is \emph{peripheral}, which is the starting point to
define Bavelas's \emph{closeness centrality} as the reciprocal of peripherality
(i.e., the reciprocal of the distances to all other nodes).

Interestingly, many of these indices
can be recast in terms of suitable calculations using the sizes
of the balls of varying radius around a node.
In a previous work~\cite{BRVH} we presented HyperANF, a tool that can compute
the distance distribution of very large graphs. HyperANF has been used, for instance, to show
that Facebook has just four ``degrees of separation"~\cite{BBRFDS}. The goal of
this paper is to extends the HyperANF approach to compute a
number of centrality indices based on distances.

Beside large-scale experiment using the full ClueWeb09 graph (almost five
billion nodes), we provide an empirical evaluation of the accuracy of our method
through a comparison with the exact centrality values on a snapshot of Wikipedia (on larger graphs the exact computation would be infeasible).
We also provide comparisons with a MapReduce-based~\cite{DeGMR} approach~\cite{KPSCLNAO}, showing that a careful
combination of HyperLogLog counters, compression and succinct data structure
can provide a speedup of two orders of magnitude, and in fact, comparing costs,
more scalability. We also show how to extend our techniques to a class of weighted
graphs with a tiny loss in space.

The Java software implementing the algorithms described in this
paper is distributed as free software within the WebGraph
framework.\footnote{\texttt{http://webgraph.di.unimi.it/}} Moreover, all
dataset we use are publicly available.

Using our Java tool we are able, for the first time, to approximate
distance-based centrality indices on graphs with billions of nodes using a standard
workstation.

\section{Notation}

In this paper, we use the following notation: $G=(V,E)$ is a directed graph with
$n=|V|$ nodes and $m=|E|$ arcs; we write $x \to y$ as a shortcut for $(x,y) \in
E$.
The length of the shortest path from $x$ to $y$ is denoted by $d(x,y)$ and
called the \emph{distance} between $x$ and $y$; we let $d(x,y)=\infty$ if there is no directed path 
from $x$ to $y$.  The nodes \emph{reachable} from $x$ are the nodes $y$ such that $d(x,y)<\infty$.
The nodes \emph{coreachable} from $x$ are the nodes $y$ such that $d(y,x)<\infty$. 
We let $G^T$ be the \emph{transpose} of $G$ (i.e., the graph obtained by reverting all arc directions in $G$).
The ball of radius $r$ around $x$ is \[\BB G(x,r)= \{\,y \mid d(x,y) \leq t\,\}.\]

\section{Geometric centralities}

We call \emph{geometric} those centrality measures\footnote{Most centrality measures proposed in the literature were
actually described only for undirected, connected graphs. Since the study of
web graphs and online social networks has posed the problem of extending
centrality concepts to networks that are directed, and possibly not strongly connected,
in the rest of this paper we consider measures depending on the \emph{incoming}
arcs of a node, so distances will be taken from all nodes
to a fixed node. If necessary, these measures can be called ``negative'', as opposed
to the ``positive'' versions obtained by taking the transpose of the graph.} whose basic assumption is that
importance depends on some function of the distances. These are actually some
of the oldest measures defined in the literature. 

\subsection{Closeness centrality}
Closeness was introduced by Bavelas in the late forties~\cite{BavCPTOG}; the closeness of $x$ is defined by 
\begin{equation}
\label{eq:closeness}
	 \frac1{\sum_y d(y,x)}.
\end{equation}
The intuition behind closeness is that nodes with a large sum of distances are
\emph{peripheral}. By reciprocating the sum, nodes with a smaller
denominator obtain a larger centrality.
We remark that for the above definition to make sense, the graph
needs be strongly connected. Lacking that condition, some of the denominators
will be $\infty$, resulting in a rank of zero for all nodes which cannot coreach the whole graph.

In fact, it was not probably in Bavelas's intentions to apply the measure to
non-connected graphs, but nonetheless the measure is sometimes ``patched'' by simply not including pairs with infinite distance, that is,
\[
\frac1{\sum_{d(y,x)<\infty} d(y,x)};
\]
for the sake of completeness, one further assumes that nodes with an empty
coreachable set have centrality $0$ by definition.
These apparently innocuous adjustments, however, introduce a strong bias toward
nodes with a small coreachable set.

\subsection{Lin's centrality}
Nan Lin~\cite{LinFSR} tried to patch the definition of closeness for graphs
with infinite distances by weighting closeness using the
square of the number of coreachable nodes; his definition for the centrality of
a node $x$ with a nonempty coreachable set is
\[
	\frac{\bigl|\{y\mid d(y,x)<\infty\}\bigr|^2}{\sum_{d(y,x)<\infty} d(y,x)}.
\]
Nodes with an empty coreachable set have centrality $1$ by definition.

The rationale behind this definitions is the following: first, we consider closeness not the inverse
of a sum of distances, but rather the inverse of the \emph{average} distance, which entails a first
multiplication by the number of coreachable nodes. This change normalizes closeness across the
graph. Now, however, we want nodes with a larger coreachable set to be more important, given
that the average distance is the same, so we multiply again by the number of coreachable nodes.

Lin's index was somewhat surprisingly ignored in the following literature.
Nonetheless, it seems to provide a reasonable solution for the problems caused by the definition of closeness.

\subsection{Harmonic centrality}
As we noticed, the main problem with closeness lies in the presence of pairs of
unreachable nodes. In~\cite{BoVAC}, we have proposed to replace the reciprocal
of the sum of distances in the definition of closeness with the sum of reciprocals
of distances. Conceptually, this corresponds to replacing the
reciprocal of a denormalized average of distances with the the reciprocal of a
denormalized \emph{harmonic} mean of distances, analogously to what Marchiori and Latora
proposed to do with the notion of average distance~\cite{MaLHSW}. The harmonic mean has the
useful property of handling $\infty$ cleanly (assuming, of course, that
$\infty^{-1}=0$).

We thus obtain the \emph{harmonic
centrality} of $x$:
\begin{equation}
\label{eq:harmonic}
	\sum_{y\neq x}\frac1{d(y,x)} = \sum_{d(y,x)<\infty, y\neq x}\frac1{d(y,x)}.
\end{equation}
The difference with~(\ref{eq:closeness}) might seem minor, but actually it
is a radical change. Harmonic centrality is strongly correlated to closeness centrality in simple networks, but naturally also accounts for 
nodes $y$ that cannot reach
$x$. Thus, it can be fruitfully applied to graphs that are not strongly connected.

\section{HyperBall}

In this section, we present \emph{HyperBall}, a general framework for
computations that depend on the number of nodes at distance at most $t$ or
exactly $t$ from a node.
HyperBall uses the same dynamic programming scheme of algorithms that
approximate neighborhood functions, such as ANF~\cite{PGFANF} or
HyperANF~\cite{BRVH}, but instead of aggregating at each step the information
about all nodes into a single output value (the neighbourhood function at $t$)
HyperBall makes it possible to perform a different set of operations (for example, depending
on the centrality to be computed). We have tried to make the treatment
self-contained, albeit a few details will be only sketched here, when they can be deduced from the description of HyperANF~\cite{BRVH}.

\subsection{HyperLogLog counters}
\label{sec:hyper}

\emph{HyperLogLog counters}, as described in~\cite{FFGH} (which is based
on~\cite{DuFLCLC}), are used to count approximately the number of distinct
elements in a stream. For the purposes of the present paper, we need to recall briefly their behaviour. 
Essentially, these probabilistic counters are a sort of \emph{approximate set
representation} to which, however, we are only allowed to pose questions about
the (approximate) size of the set.

Let $\DD$ be a fixed domain and $h: \DD \to 2^\infty$ be a fixed hash function mapping
each element of $\DD$ into an infinite binary sequence.
For a given $x \in 2^\infty$, let $h_t(x)$ denote the sequence made by the
leftmost $t$ bits of $h(x)$, and $h^t(x)$ be the sequence of remaining bits of
$x$; $h_t$ is identified with its corresponding integer
value in the range $\{\,0,1,\dots,2^t-1\,\}$. Moreover,
given a binary sequence $w$, we let $\rho^+(w)$ be the number of leading zeroes
in $w$ plus one (e.g., $\rho^+(00101)=3$). Unless
otherwise specified, all logarithms are in base 2.

% \begin{algorithm}
% \begin{tabbing}
% \setcounter{prgline}{0}
% \hspace{0.5cm} \= \hspace{0.3cm} \= \hspace{0.3cm} \= \hspace{0.3cm} \=
% \hspace{0.3cm} \= \hspace{0.3cm} \=\kill\\
% \pl\>$h: \DD \to 2^\infty$, a hash function from the domain of items\\
% \pl\>$M[-]$, an array of $m=2^b$ registers\\
% \pl\>\>(indexed from 0) and set to $-\infty$\\ \pl\>
% \FOREACH item $x$ seen in the stream \BEGIN\\ 
% \pl\>\>$i\leftarrow h_b(x)$\\
% \pl\>\>$M[i]\leftarrow\max \bigl\{M[i], \rho^+\bigl(h^b(x)\bigr)\bigr\}$\\
% \pl\>\END;\\
% \pl\>$Z\leftarrow \left(\sum_{j=0}^{m-1} 2^{-M[j]}\right)^{-1}$\\
% \pl\>return $E=\alpha_m m^2 Z$\\
% \end{tabbing}
% \caption{\label{algo:Hyperloglog}The Hyperloglog counter, as described
% in~\cite{FFGH}: it allows one to count (approximately) the number of distinct
% elements in a stream. $\alpha_m$ is a constant whose value depends on $m$ and
% is provided in~\cite{FxFGH}.}
% \end{algorithm}
% 
\begin{algorithm}
\begin{tabbing}
\setcounter{prgline}{0}
\hspace{0.5cm} \= \hspace{0.3cm} \= \hspace{0.3cm} \= \hspace{0.3cm} \=
\hspace{0.3cm} \= \hspace{0.3cm} \=\kill\\
\pl\>$h: \DD \to 2^\infty$, a hash function from the domain of items\\
\pl\>$M[-]$ the counter, an array of $p=2^b$ registers\\
\pl\>\>(indexed from 0) and set to $-\infty$\\ 
\pl\>\\
\pl\>\FUNCTION $\mathrm{add}(\text{$M$: counter},\text{$x$: item})$ \\
\pl\>\BEGIN\\
\pl\>\>$i\leftarrow h_b(x)$;\\
\pl\>\>$M[i]\leftarrow\max \bigl\{M[i], \rho^+\bigl(h^b(x)\bigr)\bigr\}$\\
\pl\>\END; \COMMENT{// function add}\\
\pl\>\\ 
\pl\>\FUNCTION $\mathrm{size}(\text{$M$: counter})$ \\
\pl\>\BEGIN\\
\pl\>\>$Z\leftarrow \left(\sum_{j=0}^{p-1} 2^{-M[j]}\right)^{-1}$;\\
\pl\>\>\RETURN $E=\alpha_p p^2 Z$\\
\pl\>\END; \COMMENT{// function size}\\
\pl\>\\ 
\pl\>\FOREACH item $x$ seen in the stream \BEGIN\\ 
\pl\>\>add($M$,$x$)\\
\pl\>\END;\\
\pl\>print $\mathrm{size}(M)$\\
\end{tabbing}
\caption{\label{algo:Hyperloglog}The Hyperloglog counter as described
in~\cite{FFGH}: it allows one to count (approximately) the number of distinct
elements in a stream. $\alpha_p$ is a constant whose value depends on $p$ and
is provided in~\cite{FFGH}. Some technical details have been simplified.}
\end{algorithm}

The value $E$ printed by Algorithm~\ref{algo:Hyperloglog}
is~\cite{FFGH}[Theo\-rem 1] an asymptotically almost\footnote{For the purposes of this paper, in the following we will
consider in practice the estimator as it if was unbiased, as suggested in~\cite{FFGH}.} unbiased estimator for the
number $n$ of distinct elements in the stream; for $n \to\infty$, the
\emph{relative standard deviation} (that is, the ratio between the standard
deviation of $E$ and $n$) is at most $\beta_p/\sqrt p\leq 1.06/\sqrt{p}$, where
$\beta_p$ is a suitable constant.
Moreover, even if the size of the registers (and of the hash function) used by
the algorithm is unbounded, one can limit it to $\log\log(n/p)+\omega(n)$ bits
obtaining almost certainly the same output ($\omega(n)$ is a function going to
infinity arbitrarily slowly); overall, the algorithm requires $(1+o(1)) \cdot p
\log\log(n/p)$ bits of space (this is the reason why these counters are called
HyperLogLog). Here and in the rest of the paper we tacitly assume that $p\geq
16$ and that registers are made of $\lceil \log\log n\rceil$ bits. 
%Note that for
%$p=16$ this implies a core-memory usage of at most $12$ bytes per counter if $n\leq 2^{64}$.

\subsection{Estimating balls}

The basic idea used by algorithms such as ANF~\cite{PGFANF} and HyperANF~\cite{BRVH}
is that that $\BB G(x,r)$, the ball of radius $r$ around node $x$, satisfies
\begin{align*}
\BB G(x,0) &= \{\,x\,\}\\
\BB G(x,r+1) &= \bigcup_{x\to y}\BB G(y,r) \cup \{\,x\,\}.
\end{align*}
We can thus compute $\BB G(x,r)$ iteratively using
sequential scans of the graph (i.e., scans in which we go in turn through the
successor list of each node). One obvious drawback of this solution is that during the scan we
need to access randomly the sets $\BB G(x,r-1)$ (the sets $\BB G(x,r)$ can be just
saved on disk on an \emph{update file} and reloaded later). For this to be possible,
we need to store the (approximated) balls in a data structure that can be fit in the
core memory: here is where
probabilistic counters come into play; to be able to use them, though, we need 
to endow counters with a primitive for the union.
Union can be implemented provided that the counter associated with
the stream of data $AB$ can be computed from the counters associated with $A$ and $B$; in the case of
HyperLogLog counters, this is easily seen to correspond to maximising the
two counters, register by register.

Algorithm~\ref{algo:DP}, named \emph{HyperBall}, describes our strategy to compute centralities.
We keep track of one HyperLogLog counter for each node; at the $t$-th iteration of the main loop, 
the counter $c[v]$ is in the same state as if it would have been fed with $\BB G(v,t)$, 
and so its expected value is $|\BB G(v,t)|$. During the execution of the loop, when we have finished
examining node $v$ the counter $a$ is in the same state as if it would have been fed with $\BB G(v,t+1)$, 
and so its value will be $|\BB G(v,t+1)|$ in expectation.

This means, in particular, that it is possible to compute an approximation of
\[|\{\, y\mid d(x,y)=t\,\}|\] (the number of nodes at distance $t$ from $x$) by
evaluating \[|\BB G(v,t+1)|-|\BB G(v,t)|.\] The computation would be exact if the
algorithm had actually kept track of the set $\BB G(x,t)$ for each node, something
that is obviously not possible; using probabilistic counters makes this feasible, at the cost of
tolerating some approximation in the computation of cardinalities.

The idea of using differences between ball sizes to estimate the number of nodes
at distance $t$ appeared also in~\cite{KTAHMRLG}, where it was used with a
different kind of counter (Martin--Flajolet) to estimate the $90\%$ percentile
of the distribution of distances from each node. An analogous technique, always
exploiting Martin--Flajolet counters, was adopted in~\cite{KPSCLNAO} to
approximate closeness. In both cases the implementations were geared towards
MapReduce~\cite{DeGMR}. A more sophisticated approach, which can be implemented
using breadth-first visits or dynamic programming, uses \emph{all-distances
sketches}~\cite{CohADSR}: it provides better error bounds, 
but it requires also significantly more memory.

\begin{algorithm}
\begin{tabbing}
\setcounter{prgline}{0}
\hspace{0.5cm} \= \hspace{0.3cm} \= \hspace{0.3cm} \= \hspace{0.3cm} \=
\hspace{0.3cm} \= \hspace{0.3cm} \=\kill\\
\pl\>$c[-]$, an array of $n$ HyperLogLog counters\\
\pl\>\\ 
\pl\>\FUNCTION $\mathrm{union}(\text{$M$: counter},\text{$N$: counter})$ \\
\pl\>\>\FOREACH $i<p$ \BEGIN\\
\pl\>\>\>$M[i] \leftarrow \max(M[i],N[i])$\\
\pl\>\>\END\\
\pl\>\END; \COMMENT{// function union}\\
\pl\>\\ 
\pl\>\FOREACH $v\in n$ \BEGIN\\
\pl\>\>$\mathrm{add}(c[v],v)$\\
\pl\>\END;\\ 	
\pl\>$t\leftarrow 0$;\\
\pl\>\DO \BEGIN\\
\pl\>\>\FOREACH $v\in n$ \BEGIN\\
\pl\>\>\>$a\leftarrow c[v]$;\\
\pl\>\>\>\FOREACH $v\rightarrow w$ \BEGIN\\
\pl\>\>\>\>$a\leftarrow \mathrm{union}(c[w],a)$\\
\pl\>\>\>\END; 	\\
\pl\>\>\>write $\langle v,a\rangle$ to disk\\
\pl\>\>\>do something with $a$ and $c[v]$\\
\pl\>\>\END; 	\\
\pl\>\>Read the pairs $\langle v,a\rangle$ and update the array $c[-]$\\
\pl\>\>$t\leftarrow t+1$\\
\pl\>\UNTIL no counter changes its value.
\end{tabbing}
\caption{\label{algo:DP}HyperBall in pseudocode. The
algorithm uses, for each node $v\in n$, an initially empty HyperLogLog counter
$c[v]$. The function $\mathrm{union}(-,-)$ maximises two counters register by
register. At line~19, one has the estimate of $|\BB G(v,t)|$ from $c[v]$ and the estimate
of $|\BB G(v,t+1)|$ from $a$.}
\end{algorithm}

HyperBall is run until 
all counters stabilise (e.g., the last iteration must leave all counters
unchanged). As shown in~\cite{BRVH}, any alternative termination condition may lead to
arbitrarily large mistakes on pathological graphs.

\section{Estimating centralities}

It should be clear that exactly three ingredients for each node $x$ are necessary to compute
closeness, harmonic, and Lin's centrality:
\begin{itemize}
  \item the sum of the distances to $x$;
  \item the sum of the reciprocals of the distances to $x$;
  \item the size of the coreachable set of $x$.
\end{itemize}
The last quantity is simply the value of each counter $c[v]$ in HyperBall at the end of
the computation on $G^T$. The other quantities can be easily computed in a cumulative fashion
nothing that 
\begin{multline*}
\sum_y d(y,x)= \sum_{t> 0}t|\{\, y\mid d(y,x)=t\,\}| \\=  \sum_{t>0}t\bigl(|\BB{G^T}(x,t)|-|\BB{G^T}(x,t-1)|\bigr),
\end{multline*} 
and
\begin{multline*}
\sum_{y\neq x} \frac1{d(y,x)}= \sum_{t> 0}\frac1t|\{\, y\mid d(y,x)=t\,\}| \\=  \sum_{t>0}\frac1t\bigl(|\BB{G^T}(x,t)|-|\BB{G^T}(x,t-1)|\bigr).
\end{multline*} 

We can thus obtain estimators for the first two ingredients by
storing a single floating point value per node, and cumulating the values for each node during
the execution of HyperBall. Note that we have to run the algorithm on the \emph{transpose}
of $G$, since we need to estimate the distances \emph{to} $x$, rather than \emph{from} $x$.

If we accept the minimum possible precision ($16$ registers per HyperLogLog counter), the core memory necessary
for running HyperBall is just 16 bytes per node (assuming $n\leq 2^{64}$), plus four booleans per node to keep track of modifications,
and ancillary data structures that are orders
of magnitude smaller. A machine with 2\,TiB of core memory could thus compute centralities on networks
with more than a hundred billion nodes, prompting the title of this paper. 
 
Note that even if we use a small number of registers per HyperLogLog counter, by
executing HyperBall multiple times we can increase the confidence in the
computed value for each estimator, leading to increasingly better
approximations.

As in the case of the average distance~\cite{BRVH}, the theoretical bounds are
quite ugly, but actually the derived values we compute are very precise, as shown
by the concentration of the values associated several runs. Multiple runs in
this case are very useful, as they make it possible to compute the empirical
standard deviation.

\subsection{Representing and scanning the graph}

In the previous section we have estimated the core memory usage of HyperBall
without taking the graph size into account. However, representing and
accessing the graph is a nontrivial problem, in particular during the last phases of the
computation, where we can keep track of the few nodes that are modifying their counter,
and propagate new values only when necessary.

Here we exploit two techniques: \emph{compression}, to represent the graph as a
bit stream in a small amount of disk space, so that we are able to access it
from disk efficiently using memory mapping; and \emph{succint data structures},
to access quickly the bitstream in a random fashion.

In particular, for compression we use the WebGraph framework~\cite{BoVWFI},
which is a set of state-of-the-art algorithms and codes to compress web and social
graphs. WebGraph represents a graph as a bitstream, with a further 64-bit pointer for each node
if random access is necessary. To store the pointers
in memory, we use a succinct encoding based on a broadword
implementation~\cite{VigBIRSQ} of the Elias-Fano
representation of monotone sequences~\cite{EliESRCASF}. This way, the cost of a pointer is logarithmic
in the average length per node of the bitstream, and in real-world graphs
this means about one byte of core memory per node, which is an order of magnitude
less than the memory used by HyperBall.

\subsection{Error bounds}

The estimate $\hBB G(x,t)$ for $|\BB G(x,t)|$ obtained by HyperBall follow the
bounds given in Section~\ref{sec:hyper}.
Nonetheless, as soon as we consider the differences $\hBB G(x,t+1)-\hBB G(x,t)$,
the bounds on the error become quite ugly. 
A similar problem occurs when estimating the distance distribution and its
statistics: by taking the difference between points of the cumulative
distribution, the bound on the relative standard deviation is lost~\cite{BRVH}.

Note that in part this is an intrinsic problem: HyperBall
essentially runs in quasi-linear expected time $O(p m \log
n)$~\cite{CohADSR}, and due to known bounds on the approximation the
diameter~\cite{RoWFAADRSG} it is unlikely that it can provide in all cases a good
approximation of the differences (which would imply a good approximation of the
eccentricity of each node, and in the end a good approximation of the diameter).

Nonetheless, for a number of reasons the estimates of the differences on real-world graphs turn out to be
very good. First of all, for very small numbers the HyperLogLog
counters compute a different estimator (not shown in
Algorithm~\ref{algo:Hyperloglog}) that is much more accurate. Second, on social
and web graphs (and in general, for small-world graphs) the function $|\BB
G(x,t)|$ grows very quickly for small values of $t$, so the magnitude of the
difference is not far from the magnitude of the ball size, which makes the relative
error on the ball size small with respect to the difference. 
Third, once most of the nodes in the reachable set are contained in $\BB G(x,t)$,
the error of the HyperLogLog counter tends to stabilise, so the bound on the
relative standard deviation ``transfers'' to the differences.

We thus expect (and observe) that the estimation of the size of the nodes at distance $t$ to be
quite accurate, in spite of the difficulty of proving a theoretical error bound.

From a practical viewpoint, the simplest way of controlling the error is
generating multiple samples, and computing the empirical standard
deviation. This is, for example, the way in which the results for the ``degrees
of separation'' in~\cite{BBRFDS} were reported. By generating several samples,
we can restrict the confidence interval for the computed values.

In Section~\ref{sec:exp} we report experiments on a relatively small graph on which
centralities could be computed exactly to show that the precision obtained on the
final values is very close to the theoretical prediction for a single counter.

\iffalse
Let $\hBB G(x,t)$ be the estimate of $|\BB G(x,t)|$ produced by Algorithm~\ref{algo:DP} at the $t$-th iteration;
then:

\begin{theorem}
For every node $x$ and every $t$
\[
	\left|\frac{E\bigl[\hBB G(x,t+1)-\hBB G(x,t)\bigr]}{|\BB G(x,t+1)-\BB G(x,t)|} -1 \right| \leq 2n\delta_1(n)+o(n)
\]
where $\delta_1$ is the same as in~\cite{FFGH}[Theorem 1] (and
$|\delta_1(x)|<5\cdot 10^{-5}$ as soon as we are using at least $16$ registers per counter).
\end{theorem} 
\begin{proof}
As in the proof of Theorem 1 of~\cite{BRVH}, we have that \[E[\hBB G(x,t)]=|\BB G(x,t)|\left(1+\delta_1(n)+o(1)\right).\]
Thus,
\begin{multline*}
	E\bigl[\hBB G(x,t+1)-\hBB G(x,t)\bigr]\\=|\BB G(x,t+1)|\left(1+\delta_1(n)+o(1)\right)+\\-|\BB G(x,t)|\left(1+\delta_1(n)+o(1)\right).
\end{multline*}
Now, dividing both members by $|\{y \mid d(x,y)=t+1\}|=|\BB G(x,t+1)|-|\BB G(x,t)|$ we obtain
\begin{multline*}
	\frac{E[\hBB G(x,t+1)-\hBB G(x,t)]}{|\{y \mid d(x,y)=t+1\}|}\\
	=1 + \frac{|\BB G(x,t+1)|\left(\delta_1(n)+o(1)\right)}{|\{y \mid d(x,y)=t+1\}|}+\\
	-\frac{|\BB G(x,t)|\left(\delta_1(n)+o(1)\right)}{|\{y \mid d(x,y)=t+1\}|}.
\end{multline*}
Since $|\BB G(x,t)|\leq n$, we obtain the result.
\end{proof}
\fi

\section{Computing with weights on the nodes}
\label{sec:nodeweights}

It is very natural, in a number of contexts, to have \emph{weights} on the
nodes that represent their importance. Centrality measures should then be redefined
taking into account weights in the obvious way: the sum of distances should
become \[\sum_{y}w(y)d(y,x),\] the sum of inverse distances should become
\[\sum_{y}\frac{w(y)}{d(y,x)},\] and the size of the coreachable set should become
\[\sum_{d(y,x)<\infty}w(y).\]

There is no direct way to incorporate weights in the dynamic programming algorithm, but weights
can be easily simulated if they are integers.
Suppose that the weighting function is $w: V \to \{1,\dots,W\}$, and assume that each node $x \in V$ is
associated with a set $\RR(x)=\{x_1,\dots,x_{w(x)}\}$ of replicas of the node (with the proviso that distinct nodes have
disjoint replicas).

Then the \emph{weighted ball of radius $r$ around $x$} can be defined recursively as:
\begin{align*}
	\WW G(x,0) &= \RR(x) \\
	\WW G(x,r+1) &= \RR(x) \cup \bigcup_{x \to y} \WW G(y,r).
\end{align*}
It is easy to see that 
\[
	|\WW G(x,r+1)|- |\WW G(x,r)|=\sum_{y: d(x,y)=r} w(y).
\]
Attention must be paid, of course, to the sizing of the counters in this case. Instead of $\log\log n$ bits,
counters with \[\log\log \sum_x w(x)\leq \log\log (Wn)=\log(\log n + \log W)\] bits will have to be used. 
We note, however, that since the increase factor
$\sum_x w(x)/n$ passes through two logarithms, it is  unlikely that more than 6 or at most
7 bits will be ever necessary. 

\section{Computing with discount functions}

If we look at harmonic centrality from a more elementary perspective, we can see that  
when measuring the centrality of a node we start by considering
its (in)degree, that is, how many neighbours it has at distance one.
Unsatisfied by this raw measure, we continue and take into consideration nodes at distance two. However,
their number is not as important as the degree, so before adding it to the
degree we \emph{discount} its importance it by $1/2$. The process continues with
nodes at distance three, discounted by $1/3$ until all coreachable nodes have been
considered.

The essence of this process is that we are counting nodes at larger and larger
distances from the target, discounting their number based on their distance.
One can generalize this idea to a \emph{family} of centrality
measures. The idea, similar to
the definition of \emph{discounted cumulative gain} in information
retrieval~\cite{JKCGBEIRT}, is that with each coreachable node we gain some
importance.
However, the importance given by the node is \emph{discounted} by a quantity
depending on the distance that, in the case of harmonic centrality, is the
reciprocal $1/d$. Another reasonable choice is a
\emph{logarithmic} discount $1/\log(d+1)$, which attenuates even more
slowly the importance of far nodes, or a \emph{quadratic} discount $1/d^2$. More generally, the centrality of $x$ based
on a non-increasing discount function $f:\mathbf N\to\mathbf R$ is \[
\sum_{d(y,x)<\infty,y\neq x}f(d(y,x)).
\]
It can be approximated by HyperBall nothing that
\begin{multline*}
\sum_{d(y,x)<\infty,y\neq x}f(d(y,x)) =\sum_{t> 0}f(t)|\{\, y\mid d(y,x)=t\,\}| 
\\=  \sum_{t>0}f(t)\bigl(|\BB{G^T}(x,t)|-|\BB{G^T}(x,t-1)|\bigr).
\end{multline*} 

We are proposing relatively mild discount functions, in contract with
the \emph{exponential} decay used, for example, in Katz's
index~\cite{KatNSIDSA}. This is perfectly reasonable, since Katz's index is
based on \emph{paths}, which are usually infinite. Discount-based centralities
are necessarily given by finite summations, so there is no need for a rapid
decay.
Actually, by choosing a constant discount function we would estimate the
importance of each node just by the number of nodes it can coreach (i.e., in the
undirected case, by the size of its connected component).

Combining this observation and that of Section~\ref{sec:nodeweights}, we
conclude that HyperBall can compute a class of centralities that could be called
\emph{discounted-gain centralities}:\footnote{These
are called \emph{spatially decaying} in~\cite{CohSDAN}.}
\[
\sum_{d(y,x)<\infty,y\neq x}w(y)f(d(y,x)).
\]

\section{Experiments}
\label{sec:exp}

We decided to perform three kinds of experiments:
\begin{itemize}
\item A small-scale experiment on the same graphs for which explicit timings are reported in~\cite{KPSCLNAO}, to compare
 the absolute speed of a MapReduced-based approach using the Hadoop open-source implementation and of an in-core approach. Note that the graphs involved are extremely unrealistic
 (e.g., they have all diameter 2 and are orders of magnitude denser than typical web or social graphs).
 This experiment was run using $p=64$ registers per HyperLogLog counter, corresponding to a relative standard deviation of
$13.18$\%, which is slightly better than the one used in~\cite{KPSCLNAO} ($13.78$\%, as communicated by the authors), 
to make a comparison of the execution times possible.
\item A medium-size experiment to verify the convergence properties of our computations. For this purpose, we had to restrict ourselves to a
graph for which exact values could be computed using $n$ breadth-first visits. We focused on a public snapshot of Wikipedia\footnote{Available at \texttt{http://law.di.unimi.it/}}. 
This graph consists of $4\,206\,785$ nodes and $101\,355\,853$ arcs (with average degree $24$ and the largest strongly connected component spanning about 89\% of
the nodes).
We performed 100 computations using $p=4096$ registers per counters, corresponding to a theoretical relative standard deviation of $1.62\%$ for each computation.
The exact computation of the centralities required a few days using 40 cores.
\item A large-scale experiment using the largest ClueWeb09\footnote{A dataset gathered in 2009 within the U.S.~National Science Foundation's Cluster Exploratory (CluE) 
program. The ClueWeb12 graph will be even larger, but it is presently still under construction. See \url{http://lemurproject.org/clueweb09/}} graph; ClueWeb09 is, at the
time of this writing, the largest web graph publicly available, one order of magnitude larger that previous efforts in terms of nodes. It contains $4\,780\,950\,903$ nodes and $7\,939\,647\,896$ arcs.
The purpose of this experiment was to show our methods in action on a very large dataset.\footnote{We remark that due to the way in which the graph has been collected (e.g., probably starting from a large seed) the graph is actually
significantly less dense than a web graph obtained by breadth-first sampling or similar techniques. Moreover, the graph
contains the whole set of discovered nodes, even if only about $1.2$ billion pages were actually crawled. As a result, 
many statistics are off scale: the harmonic diameter~\cite{MaLHSW,BoVFDSR} is $\approx 15131$ (typical values for breadth-first 
web snapshots are $\approx 20$) and the giant component is just $0.6$\% of the whole graph.}
\end{itemize}

\begin{figure*}
\centering
\begin{tabular}{cc}
\includegraphics[scale=.45]{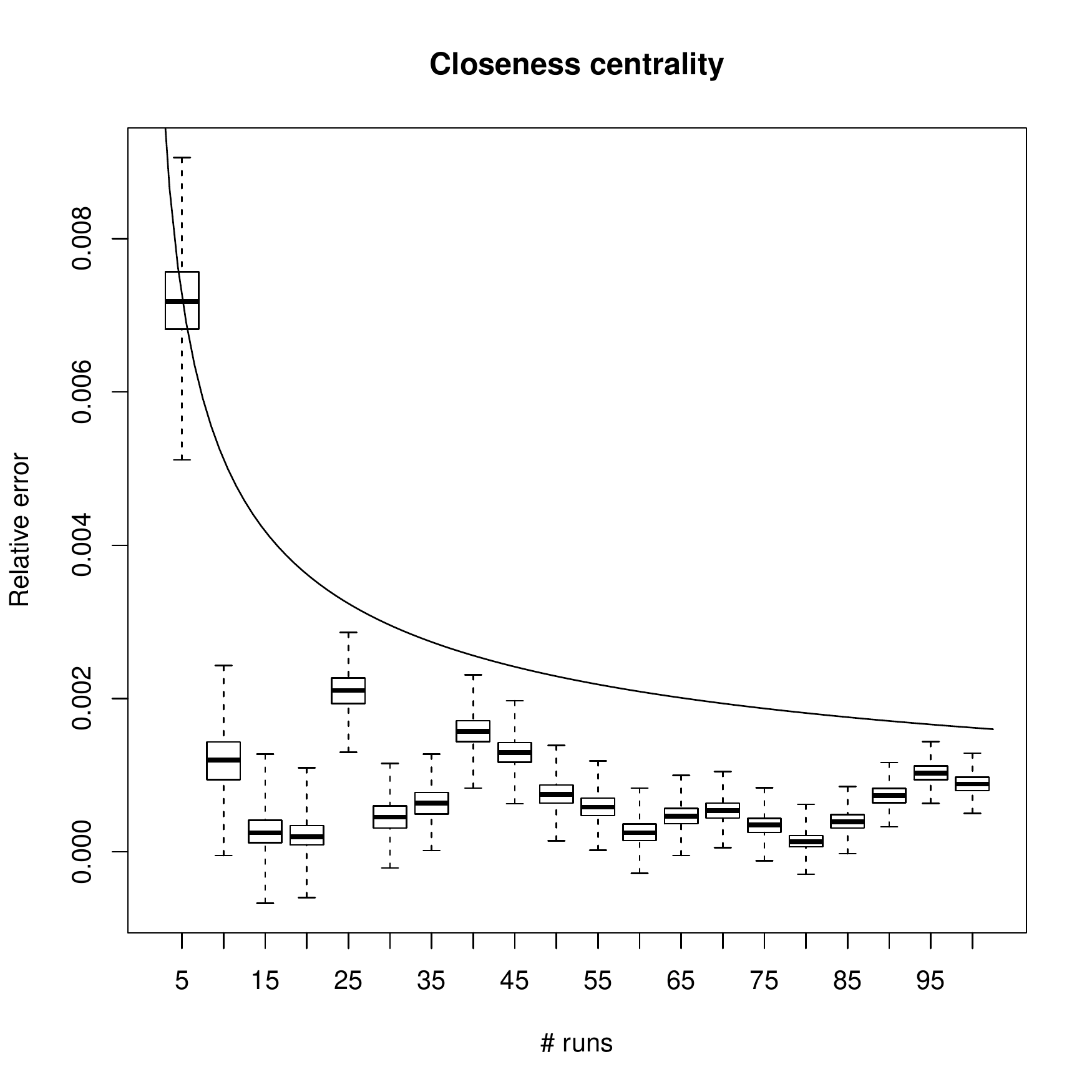} &
\includegraphics[scale=.45]{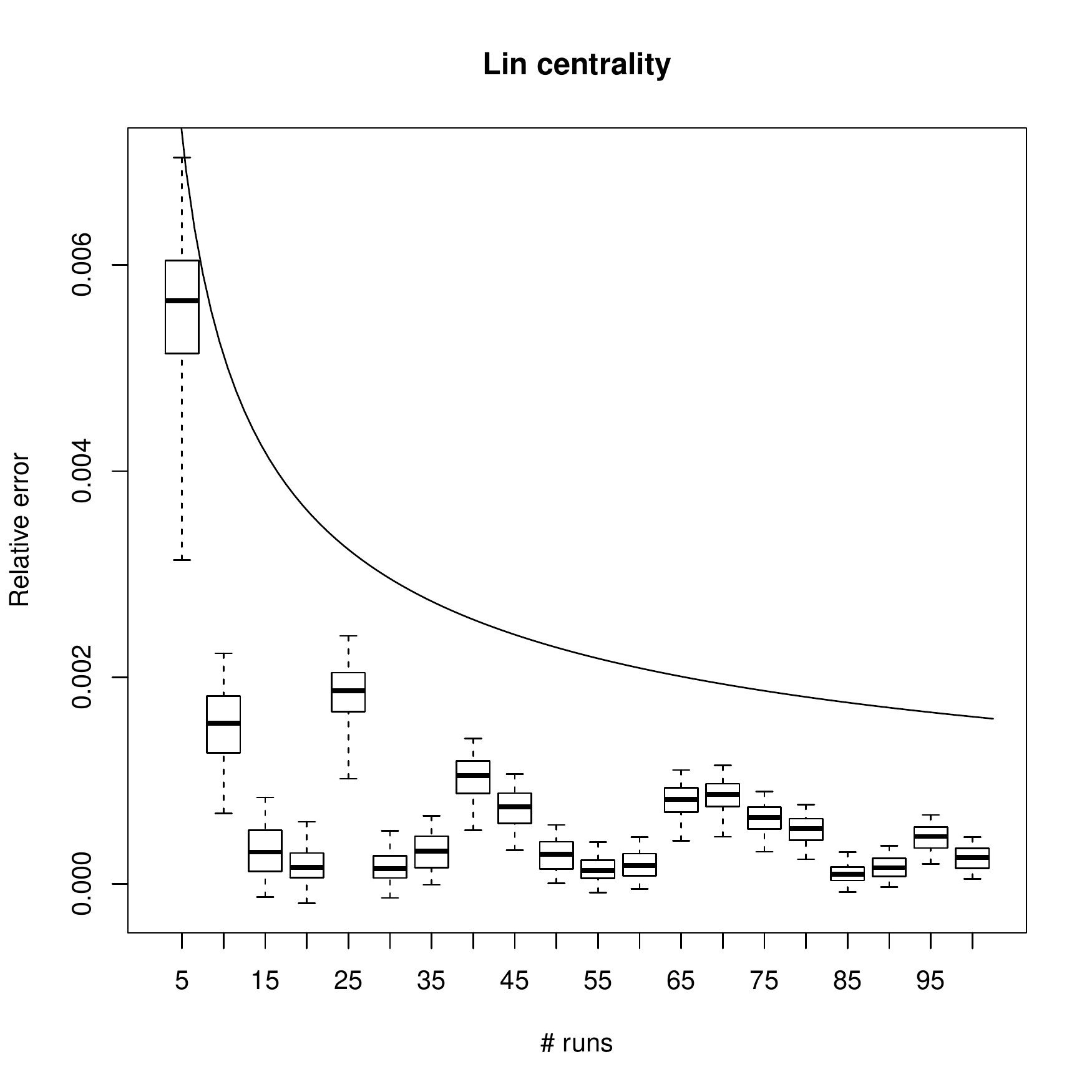}\\
\includegraphics[scale=.45]{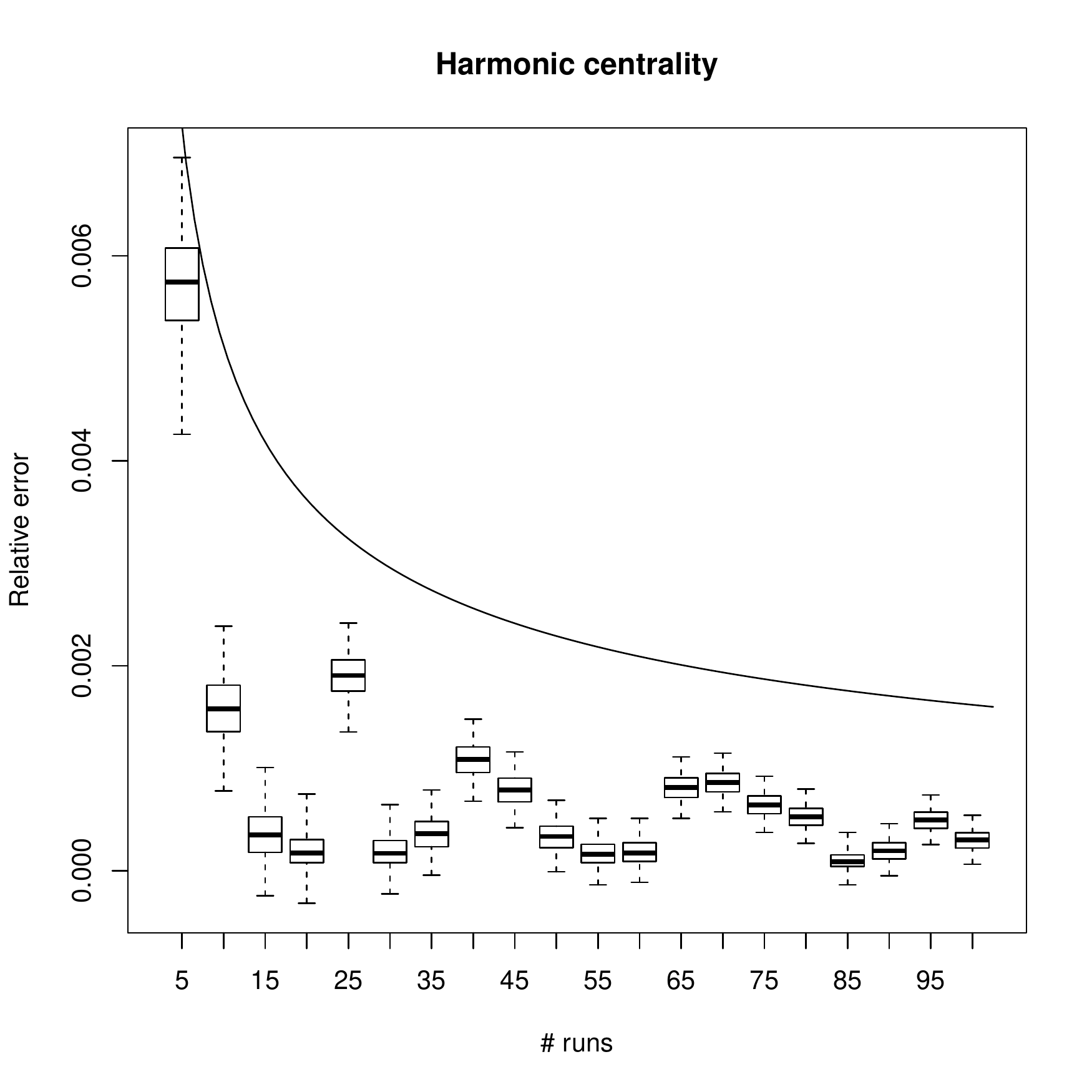}&
\includegraphics[scale=.45]{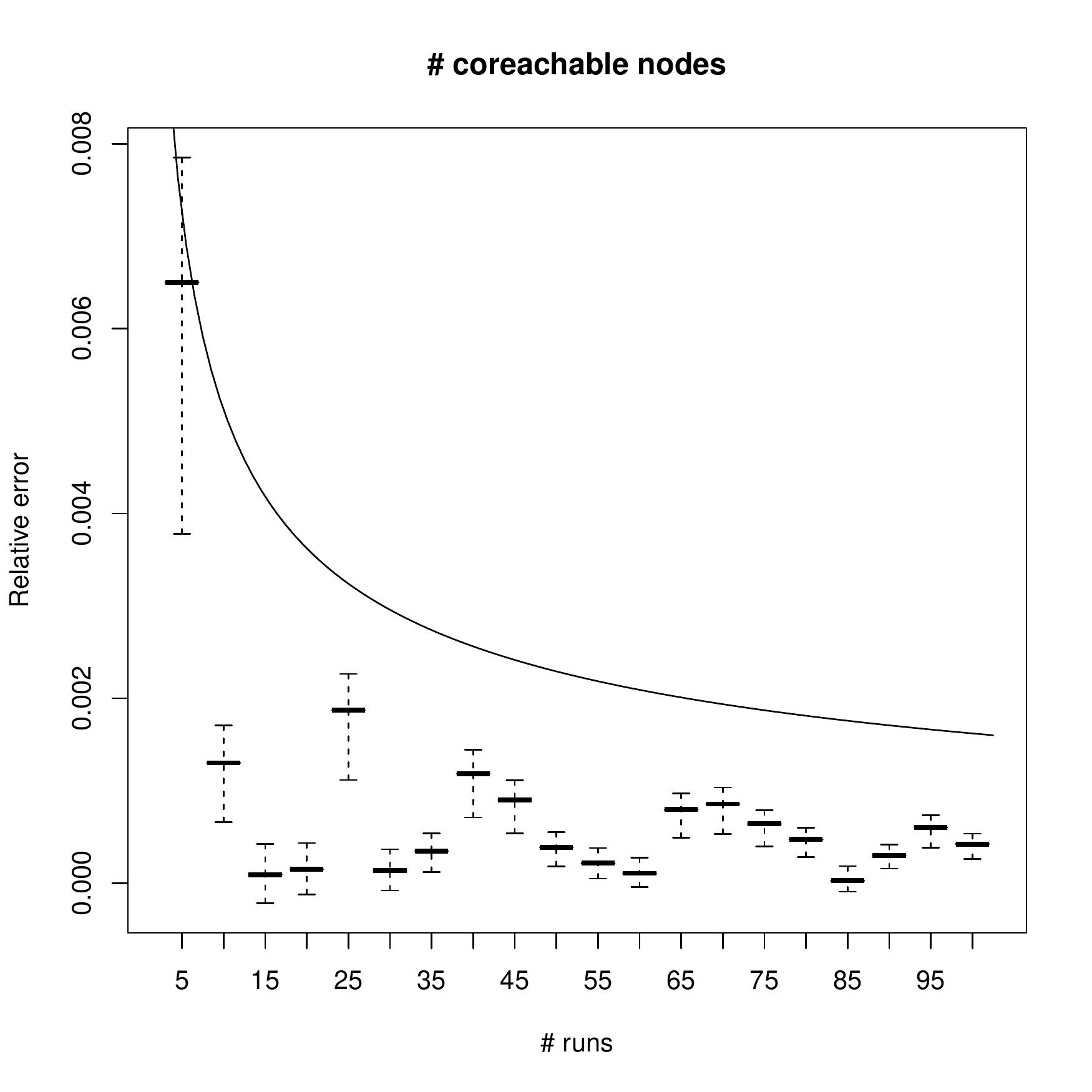}
\end{tabular}
\caption{\label{fig:enwiki} Relative errors in the computation of centrality measures on Wikipedia: we averaged the values computed
in 5, 10, 15, \dots, 100 runs and computed the relative error with respect to the real value (the latter were obtained by running an exact implementation). 
The boxes represent the 1st (lower edge), 2nd (i.e., the median; midline) and 3rd (upper edge) quartile; the whiskers correspond
to an interval of length $2\sigma$ around the mean. For comparison, each plot contains the curve of the theoretical relative standard deviation for each single HyperLogLog counter over the
given number of samples.}
\end{figure*}

\begin{table}
\caption{\label{tab:time}Comparative timings per iteration between the Hadoop implementation described in~\cite{KPSCLNAO} running on 50 machines and
HyperBall on a MacBook Pro laptop ($2.6$\,GHz Intel i7, 8\,GiB RAM, 8 cores) and on a 32-core, 64\,GiB RAM workstation using
$2.3$\,GHz AMD Opteron 6276 processors. Timings for the Hadoop implementation were deducted from Figure~4(b) of~\cite{KPSCLNAO}. Note that
the better processor and the SSD disk of the MacBook Pro make it almost twice faster (per core) than the workstation.}
\centering
\begin{tabular}{r|r|r|r}
\multicolumn{1}{c|}{Size (nodes/arcs)} & \multicolumn{1}{c|}{Hadoop~\cite{KPSCLNAO}} & \multicolumn{1}{c|}{MacBook} & \multicolumn{1}{c}{32 cores}\\
\hline
20\,K / 40\,M & $250$\,s & $2$\,s & $1$\,s\\
59\,K / 282\,M & $1750$\,s & $10$\,s & $4$\,s\\
177\,K / 1977\,M & $2875$\,s & $70$\,s & $23$\,s  
\end{tabular}
\
\end{table}

In Table~\ref{tab:time} we report the timings for an iteration on the same set
of Kronecker graphs used in~\cite{KPSCLNAO}. A standard workstation with 32 cores using HyperBall is at least 150 times faster
than a Hadoop-based implementation using using 50 machines; even a MacBook Pro with 8 cores is at least 50 times faster.

In Figure~\ref{fig:enwiki} we report the results of the second set of
experiments, which fully confirm our empirical observations on the
behaviour of the difference estimator: on average, the relative error on the computed centrality indices is very close to
the theoretical prediction for each single HyperLogLog counter, and, in fact, almost always significantly smaller.

It is interesting to observe that the estimation on the number of coreachable
nodes (depending on the value of a single counter at the end of the computation) is extremely
more concentrated. This is due both to the lack of differences, which reduces
the error, and to the fact that most nodes ($89\%$) lie in the giant strongly
connected component, so their coreachable set is identical, and this induces a
collapse of the quartiles of the error on the median value.

On the same dataset, Table~\ref{tab:scalability} reports figures showing that
increasing the number of cores leaves essentially unmodified the time per arc
per core (i.e., linear scalability). The only significant (30\%) increase happen at 32 cores, and it is
likely to be caused by the nonlinear cost of caching.

Finally, we ran HyperBall on ClueWeb09 using a
workstation with 40 Intel Xeon E7-4870 at $2.40$\,GHz and 1\,TiB of RAM (with the same hardware, we could have
analysed a graph with 50 billion nodes using $p=16$). We report the results in Table~\ref{tab:time2}. We performed
three experiments with different levels of precision, and in the one with the highest precision we fully utilized the in-core memory: 
the timings show that increasing the precision scales even better than linearly, which is to be expected, because the cost of
scanning the graph is constant whereas the cost of computing with greater precision grows linearly with the number of registers
per HyperLogLog counter. Thus, for a fixed desired precision a greater amount of in-core memory translates into higher speed.

\begin{table}
\caption{\label{tab:scalability}Time per arc per core of a HyperBall iteration, tested on the Wikipedia graph
with $p=4096$.}
\centering
\begin{tabular}{r|r|r|r}
\multicolumn{1}{c|}{cores} & \multicolumn{1}{c|}{Time per arc per core} \\
\hline
$1$ & $906$\,ns \\
$2$ & $933$\,ns \\
$4$ & $967$\,ns \\ 
$8$ & $1018$\,ns \\ 
$16$ & $1093$\,ns \\ 
$32$ & $1389$\,ns \\ 
\end{tabular}
\end{table}

\begin{table}
\caption{\label{tab:time2}Timings for a full 40-core computation ($\approx 200$ iterations) on ClueWeb09
using a different number $p$ of registers per HyperLogLog counter. The amount of memory
does not include $7.2$\,GiB of succinct data structures that store pointers to the memory-mapped
on-disk bitstreams representing the graph and its transpose.}
\centering
\begin{tabular}{l|r|r|r}
\multicolumn{1}{c|}{$p$} & \multicolumn{1}{c|}{Memory} &\multicolumn{1}{c|}{Overall time} &\multicolumn{1}{c}{Per iteration (avg.)} \\
\hline
$16$ & $73$\,GiB & 96\,m & 27\,s \\
$64$ & $234$\,GiB & 141\,m & 40\,s\\
%$128$ & $447$\,GiB & 179\,m & 50\,s\\
$256$ & $875$\,GiB & 422\,m & 120\,s \\ 
\end{tabular}
\end{table}

\section{Conclusions and future work}

We have described HyperBall, a framework for in-core approximate computation of centralities based on the
number of (possibly weighted) nodes at distance exactly $t$ or at most $t$ from
each node $x$ of a graph. With 2\,TiB of memory, HyperBall makes it
possible to compute accurately and quickly harmonic centrality for graphs up to a hundred
billion nodes. We obtain our results with a mix of approximate set
representations (by HyperLogLog counters), efficient compressed graph
handling, and succinct data structures to represent
pointers (that make it possible to access quickly the memory-mapped graph representation).

We provide experiments on a $4.8$ billion node dataset, which should be contrasted
with previous literature: the largest dataset in \cite{KPSCLNAO} contains $25$ million nodes,
and the dataset of~\cite{KTAHMRLG} contains $1.4$ billion nodes.
% \footnote{This is the Altavista
% webpage connectivity set distributed by Yahoo! as part of the WebScope program. It should be
% remarked by this graph, albeit widely used in the literature, is not a good
% dataset. The the giant component is less than 4\% of the whole graph, and 49\% of the nodes are isolated.}
Moreover, both papers provide timings only for a
small, $\approx 177\,000$-nodes graph, whereas we report timings for all our datasets.

\bibliography{biblio,harmonic,related}

\end{document}